\documentclass[traditabstract]{aa} 
\usepackage{graphicx,longtable,cancel}
\usepackage{txfonts}
\begin{document}

\bibliographystyle{aa}
\newcommand{\um}{$\mu$m}
\newcommand{\uJy}{$\mu$Jy}
\newcommand{\ergs}{{erg\,s$^{-1}$}}
\newcommand{\ergscm}{{erg\,s$^{-1}$\,cm$^{-2}$}}
\newcommand{\kms}{{km\,s$^{-1}$}}
\newcommand{\msun}{${\rm M}_{\odot}$}
\newcommand{\lsun}{${\rm L}_{\odot}$}
\newcommand{\msunyr}{${\rm M}_{\odot}~{\rm yr}^{-1}$}
\newcommand{\lir}{$L_{\rm IR}$}
\newcommand{\Lbol}{$L_{\rm bol}$}
\newcommand{\ms}{$M_{\rm star}$}
\newcommand{\mg}{$M_{\rm gas}$}
\newcommand{\mh}{$M_{\rm halo}$}
\newcommand{\mbh}{$M_{\rm BH}$}
\newcommand{\mui}{$\mu^{-1}$}
\newcommand{\cohh}{CO$\to$H$_2$}
\def\Zsun{Z_\odot}
\def\gtsima{$\; \buildrel > \over \sim \;$} 
\def\ltsima{$\; \buildrel < \over \sim \;$} \def\prosima{$\; \buildrel \propto \over \sim \;$} 
\def\gsim{\lower.5ex\hbox{\gtsima}}
\def\simgt{\lower.5ex\hbox{\gtsima}} 
\def\simlt{\lower.5ex\hbox{\ltsima}} 
\def\lsim{\lower.5ex\hbox{\ltsima}}
\def\msole{M_\odot}
\def\msun{M_\odot}
\def\OVI{\hbox{O~$\scriptstyle\rm VI$}}
\def\invMpch {~h~{\rm Mpc}^{-1}}
\def\Mpch {~h^{-1}~{\rm Mpc}}
\def\kpch {~h^{-1}~{\rm kpc}}
\def\swift {{\it Swift}}
\def\xmm {{\it XMM-Newton}}
\def\chandra {{\it Chandra}}

 \title{Missing cosmic metals revealed by X--ray absorption towards distant sources}

\author{
Sergio Campana\inst{1}, 
Ruben Salvaterra\inst{2},  
Andrea Ferrara\inst{3,4},
Andrea Pallottini\inst{3}
}

\institute{INAF, Osservatorio Astronomico di Brera, Via E. Bianchi 46, I-23807 Merate (LC), Italy\\
\email{sergio.campana@brera.inaf.it}
            \and
INAF, IASF Milano, Via E. Bassini 15, I-20133 Milano, Italy
             \and 
Scuola Normale Superiore, Piazza dei Cavalieri 7, I-56126 Pisa, Italy
             \and
Kavli IPMU (WPI), Todai Institutes for Advanced Study, University of Tokyo
        }             

   \date{ }

\abstract{
The census of heavy elements (metals) produced by all stars through cosmic times up to present-day is limited to 
$\lsim 50\%$; of these only half are still found within their parent galaxy. 
The majority of metals is expelled from galaxies into the circumgalactic (or even more 
distant, intergalactic) space by powerful galactic winds, 
leaving unpleasant uncertainty on the amount, thermal properties and distribution of these key chemical species. 
These dispersed metals unavoidably absorb soft X--ray photons from distant sources.  We show that their integrated 
contribution can be detected in the form of increasing X--ray absorption with distance, for all kinds of 
high-energy cosmic sources. Based on extensive cosmological 
simulations, we assess that $\sim$ 10\% of all cosmic metals reside in the intergalactic medium. 
Most of the X--ray absorption arises instead from a few discrete structures along the line of sight. These 
extended structures, possibly pin-pointing galaxy groups, contain million degree, metal-enriched gas, 
100--1,000 times denser than the cosmic mean. An additional $\sim 10\%$ of cosmic metals could reside in this phase.
}

\keywords{intergalactic medium, X-rays: general, gamma-ray burst: general, galaxies: active, galaxies: halos}
               
\authorrunning{Campana, Salvaterra, Ferrara \& Pallottini} 

\titlerunning{Missing cosmic metals revealed by X--ray absorption}

\maketitle

\section{Introduction}

Elements heavier than boron (in short, ``metals'') are produced only by stars and supernovae. Given the star formation history of the 
universe (Madau \& Dickinson 2014) one can compute the amount of metals produced and then dispersed in the surroundings by supernova
(SN) explosions (core-collapse SNe provide $\sim 80\%$ of the metals by mass) and stellar winds from asymptotic giant branch stars. 
The relative distribution of metals among different galactic components (galaxy, halo, intergalactic medium [IGM]) is a sensitive 
diagnostic of such complex phenomena  as star formation, galactic outflows, and galaxy interactions that might strip metals from their formation sites. 

Metal reservoirs within galaxies are stars, interstellar medium, and dust (Fukugita, Hogan \& Peebles 1998). 
At the present epoch (redshift $z=0$) the fraction of metals 
produced by star-forming galaxies and retained within them is $\sim 20-25\%$ (Peeples et al. 2014; Shull, Smith \& Danforth 2012).
This calls for an efficient transport mechanism that pours metals from galaxies into the circumgalactic medium (a few hundreds kpc size 
environment  also often referred to as the galactic halo) or the ``true'' IGM. These galactic outflows can be powered either by 
SN-driven galactic winds (Oppenheimer  \& Dav\'e 2006; Kobayashi, Springel \& White 2007) or by energy injection from an active galactic 
nucleus (Planelles et al. 2014; Barai et al. 2013).
This energy deposition may also strongly suppress the galaxy star formation activity.

Galactic haloes are characterised by a wide range of densities and temperatures. Metals in this medium are mainly detected 
through optical/UV absorption lines of intervening absorbers along the line of sight (LOS) to bright sources, typically quasars.  
A considerable observational effort has been put into these studies. The most prominent absorption features correspond to high-ionization 
species (indicating the presence of warm gas with $T\sim 10^5$ K) like O\,VI, C\,IV, Si\,IV (Tumlinson et al. 2011; Danforth et al. 2014; Werk et al. 2014), and 
dust (M\'enard et al. 2010; M\'enard \& Fukugita 2012). These metals, which can be traced up to galactocentric distances of $\sim 150-500$ kpc, add 
$\sim 20-30\%$ to the total metal census, still leaving about 50\% of the total metal budget undetected. This is known as 
the missing metals problem (Pettini et al. 1999; Ferrara, Scannapieco \& Bergeron 2005; Bouch\'e, Lehnert, P\'eroux 2005;
Pagel 2008; Ford et al. 2014).

The missing metals problem looks somewhat different at different redshifts.
The metal census at $z\sim 2$ has been analysed in a series of papers by Bouch\'e et al. (2005, 2006, 2007).
Currently known galaxy populations at $z\sim 2$ can account for $30-60\%$ of the expected metals.
Absorption line studies trace the metal content in lower density gas phases. 
Metals in the Ly$\alpha$ forest ($\log{N_{\rm HI}} < 17$) account for $\sim 15-30\%$ of the metals. 
Lyman limit systems ($\log{N_{\rm HI}} > 17$, including Damped Ly$\alpha$ systems), provide $\sim 5-15\%$ (Lehner et al. 2014; 
Bouch\'e et al. 2007) and at least $10\%$ is in the form of highly ionised metals in the circumgalactic gas of galaxies (Lehner et al. 2014). 
Summing up these contributions we conclude that $50\%$ of metals lies outside galaxies.
Based on these estimates, one can conclude that the metal census at this redshift might be almost complete. 
On the contrary, in the local universe, Peeples et al. (2014) showed that $\sim 50-55\%$ of metals can be accounted 
for within galaxies, with an error budget of $\sim \pm25\%$.
The remaining metals should be dispersed outside galaxies on a larger scale.

\begin{table*}
\begin{center}
\begin{tabular}{lcccc}
\hline
\hline
Sample         & Instrument & Detections (UL) &  Redshift range& Refs.\\
\hline
GRB-C1012& \swift\           & 107 (11)              & 0.03--8.2            & 1,2\\
AGN-E13     & \xmm\           & 24 (34)                & 0.45--4.72          &3\\
AGN-C11     & \xmm\           & 40 (245)              & 0.02--2.39          & 4\\   
AGN-Y09     & \xmm\           & 32 (141)              & 0.22--5.41          & 5\\
AGN-hz        & {\it XMM} \&  {\it Chandra} & 1 (15)  &3.63--6.28   & 6,7\\
AGN-P05     & \xmm\           & 19 (44)                & 0.01--6.28           & 8\\
\hline
\hline
\end{tabular}
\end{center}
\caption{Summary of X--ray source samples used in the text.}
\medskip
References: 1: Campana et al. (2010); 2: Campana et al. (2012); 3: Eitan \& Behar (2013); 
4: Corral et al. (2011); 5: Young, Elvis \& Risaliti (2009);  6: Vignali et al. (2003); 
7: Saez et al. (2011); 8: Page et al. (2005). 
Superposition of data were checked and older data were removed.
\label{tab:nh}
\end{table*}

Thus we are forced to make the hypothesis that the missing metals should be found at even larger distances, in the IGM or
 in the gas in between galaxies in groups and clusters. Among these possibilities, we can safely exclude only clusters, as their metal content is relatively easy to 
measure and largely insufficient to close the gap (Simionescu et al. 2011; Werner et al. 2013). 
Observations of galaxy groups, although much more difficult, 
tentatively indicate a larger metal content than the sum of single galaxies (Sun et al. 2009; Humphrey et al. 2012).
Finally, little can be said about the amount of metals in the IGM. 

Metals absorb soft X--ray photons. This ``metal fog"  should manifest itself as an increase in X--ray absorption with distance for all kinds of 
cosmic high-energy sources.  A first hint of the existence of this metal fog came from gamma-ray bursts (GRB) observations
(Campana et al. 2010; Behar et al. 2011). 
These sources, spanning a very large redshift range, are the primary probe for this effect. Despite the intrinsic absorption by gas within 
the GRB host galaxy itself (see Fig. 1), a minimum absorbing column density envelope enclosing all GRBs and increasing with redshift exists. 
If it arises from a cosmological gas distribution, this minimum cumulative absorption could affect all kinds of high-energy sources. Indeed, this effect persists when GRB data is combined with active galactic nuclei (AGN) samples(Starling et al. 2013; Eitan \& Behar 2013). 
At variance with other techniques (e.g. UV absorption line studies), X--ray absorption is less sensitive to the ionisation state and 
thermodynamic phase (gaseous vs. solid, i.e. dust) of the absorbing medium,
allowing for, in principle, an almost complete metal census. 

In this paper we explore the minimum X--ray absorbing column density based on samples of GRBs and AGNs  discussed in Section 2
(particular care is taken to discuss a few outliers present in these samples and discussed in Appendix A). 
In Section 3 we work out the lower envelope of the column density distribution as well as the mean absorbing value.
We then investigate how this absorption arises by means of cosmological simulations. In Section 4 we briefly describe our simulations. In Section 5 
we describe in detail how we evaluate the absorbing cross section of the intervening material. The evaluation of the fraction of metals in the IGM 
and in the cosmic structures is carried out in Section 6.
Discussion and conclusions are drawn in Section 7.

\section{Sample of X--ray sources} \label{sec:sample}

Our sample of absorbed X--ray sources consists of GRB and AGN.
We took data   from the literature without further intervention (expect for a few outliers, see the Appendix).
We selected data  in order to have a sufficiently large number of counts to provide a reliable measure of 
the X--ray spectrum and of the intrinsic absorption column density at the source redshift ($N_{\rm H}(z)$) in particular. 
The GRB sample relies on \swift\ data and the AGN samples on \xmm\ (see Table 1).

\begin{figure*}[!ht]
\centerline{
\includegraphics[width=1.0\textwidth]{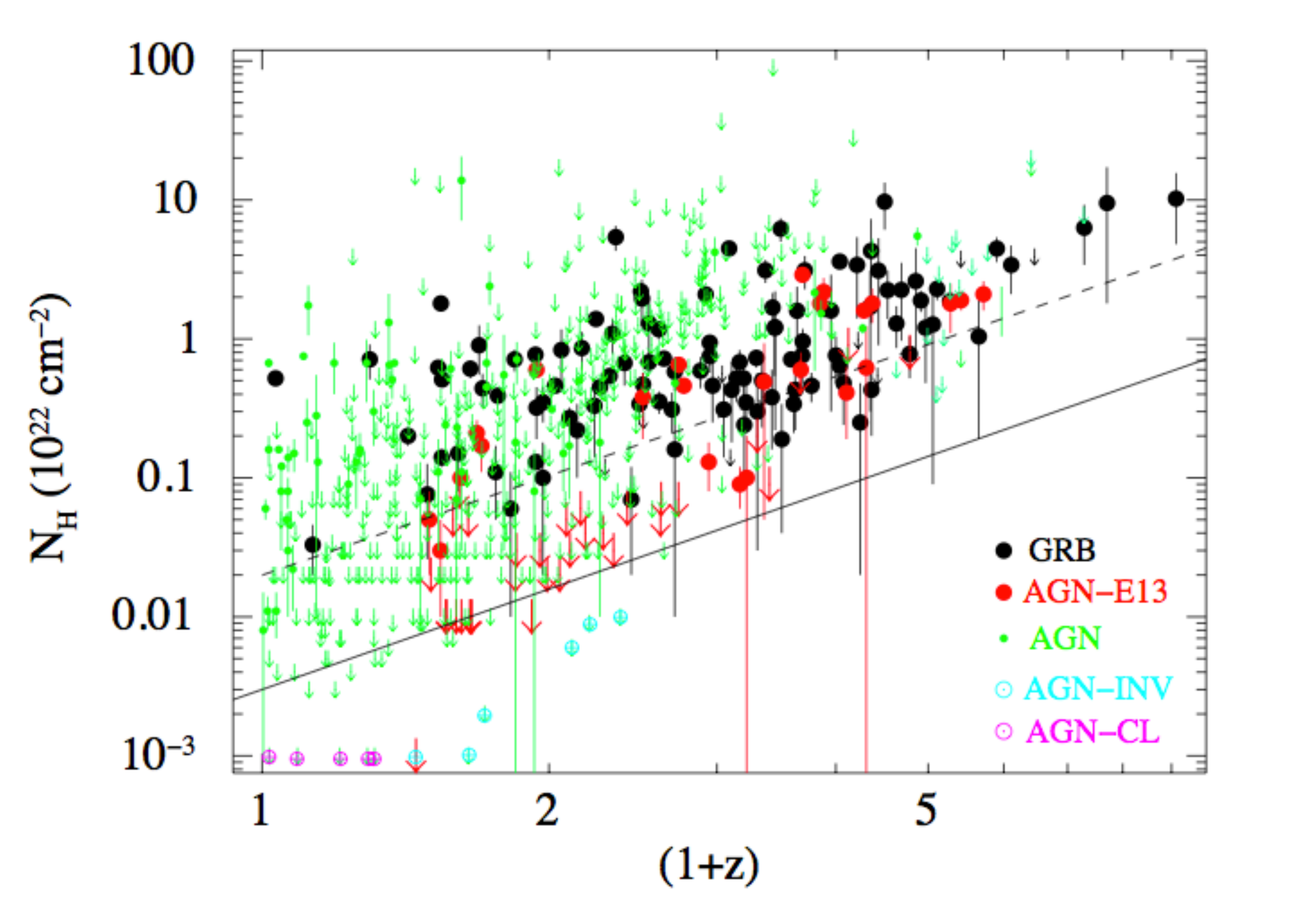}
}
\caption{Column densities of GRB and AGN of our sample as a function of redshift ($1+z$). Black symbols are used for GRB (filled circles 
for detections and arrows for upper limits). Filled red circles and arrows identify detected (upper limits) intrinsic column densities within the AGN-E13 sample,
which has been used to compute the mean absorbing column density. 
Green symbols (filled circles and arrows) identify (detections and upper limits, respectively)  AGN of the other samples. 
Circled crosses indicate AGN that were discarded from the overall sample: magenta symbols identify $z<0.4$ AGN, light blue symbols were further 
investigated in the Appendix.
The continuous line marks the lower envelope of the intrinsic column densities of $\log N_{\rm H} \simeq 19.5$ and the dashed line the mean of the 
AGN-E13 sample, as derived from the censored analysis, with $\log N_{\rm H}\simeq 20.3$.}
\label{F1}
\end{figure*}

\subsection{Gamma--ray bursts}\label{subsec:grb}

We constructed the GRB sample (GRB-C1012)  by merging the sample compiled by Campana et al. (2010) with all GRB with known redshifts
at that time together with the additional bursts quoted in Campana et al. (2012), part of the complete sample of bright GRB (Salvaterra et al. 2012).
This sample is selected based on a cut at the burst peak flux (BAT6) and is free of bias in the column density measurements, which is almost complete in redshift.
The total number of GRB amounts to 118.  We have an upper limit on the $N_{\rm H}(z)$ for only 11 of them, testifying that intrinsic absorption is 
a common characteristic of these sources.
It is apparent that higher-redshift GRBs are characterised by a larger $N_{\rm H}(z)$. 
Based on the Campana et al. (2010) sample, a Kolmogorov-Smirnov (KS) test shows the probability that the column density distributions 
of GRBs above and below $z=1$ are not drawn from the same parent distribution is 0.08 per cent (equivalent to Gaussian $3.3\,\sigma$). This supports a cosmological increase of the absorbing column density (Campana et al. 2010; Behar et al. 2011; Starling et al. 2013).

\subsection{Active galactic nuclei}\label{subsec:grb}
 AGN data were taken from selected bright samples of objects:

{\bf AGN-E13}: Eitan \& Behar (2013) constructed a sample of 58 high-redshift ($0.45\le z\le 4.72$) AGN taken from the \xmm\ archive and
selected to have $>1,800$ counts. This ensures an accurate determination of the intrinsic column density. The sample includes 
radio-quiet and radio-loud quasars (30 and 28, respectively). The sample contains 24 detections and 34 upper limits.

{\bf AGN-C11}: Corral et al. (2011) selected AGN from the \xmm\ bright survey (XBS). The XBS is composed of two flux limited 
samples selected in the medium (0.5--4.5 keV) and hard (4.5--7.5 keV) energy bands. The sample contains 270 Type I (unabsorbed) AGN, 
29 Type II (absorbed) AGN and five blazars. The distribution in redshift covers the range 0.02--2.39. 
The sample contains 40 determinations of $N_{\rm H}(z)$ and 245 upper limits.

{\bf AGN-Y09}:  Young, Elvis \& Risaliti (2009) built an AGN sample matching quasars from the $5^{\rm th}$ Sloan Digital sky survey release
with serendipitous \xmm\ observations. The initial sample comprised 685 X--ray source detections, with 589 radio-quiet, 62 radio-loud, and 34 broad 
absorption line (BAL) quasars.   A meaningful $N_{\rm H}(z)$ can be derived for only a minority of them. The final sample consists of 32 detections and 
141 upper limits. The redshift range sampled by these AGN is 0.22--5.41.

{\bf AGN-hz}: In order to increase the statistics at high redshift, we merged the small groups of sources provided by Vignali et al. (2003) and 
Saez et al. (2011).
This is based on \xmm\ and \chandra\ observations. In this sample, spanning $3.63\le z\le 6.28$, we have just one $N_{\rm H}(z)$ determination and 15 upper limits.

{\bf AGN-P05}: Page et al. (2005) selected a sample of 29 high-redshift ($z>2$) AGN, including 7 radio-quiet,
16 radio-loud, and 6 BAL quasars, and complemented it with a number of low redshift AGN (39). There is a strong overlap with previous samples and 
only 59 AGN were retained ($0.01\le z \le 4.41$). Of those, 19 have a determination of $N_{\rm H}(z)$ and 44 only an upper limit.
 
 \begin{figure}[!t]
\centerline{
\includegraphics[width=0.48\textwidth]{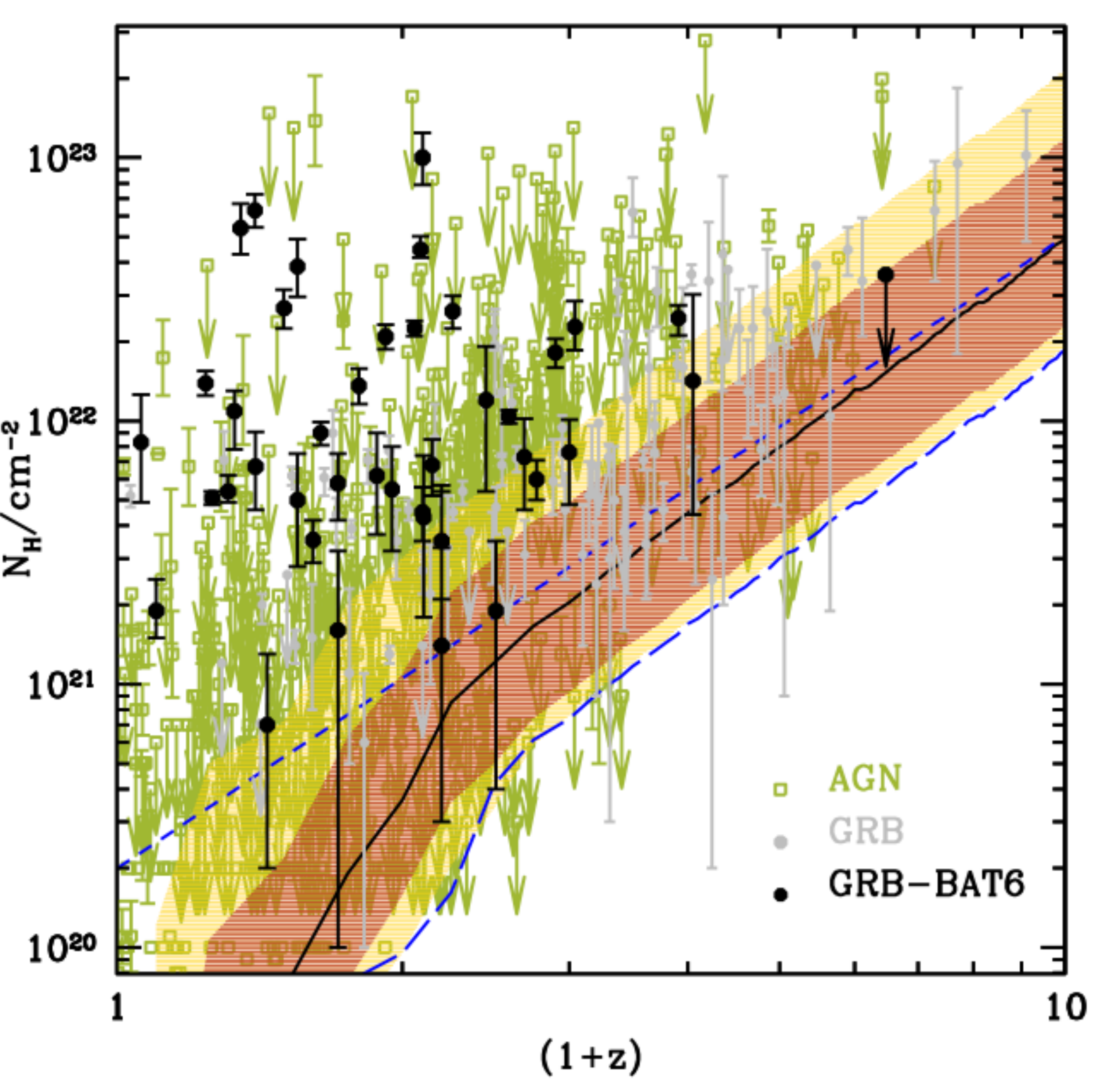}
}
\caption{Distribution of equivalent hydrogen column densities, $N_{\rm H}$, with redshift as measured for different sources: 
GRBs with measured redshift (grey dots), bright GRBs in a redshift complete sample (black dots), 
and AGN (green squares). Error bars are at $1\,\sigma$ confidence level.
The long dashed curves represents the LOS with the lowest absorption, i.e. with no absorbers with density contrast $\Delta >100$, and constant 
metallicity ($Z\sim 0.03\,\Zsun$). This curve was obtained from cosmological simulations, summing up the contribution by each cell, with an 
absorbing column density weighted for its effective (temperature-dependent) value. 
The results for a physically motivated model of SN winds (momentum-driven 
model by Oppenheimer et al. 2012) 
are shown by the solid line (median of LOS distribution) and corresponding shaded areas ($68\%$ and $90\%$ of LOS, respectively).
The short dashed line corresponds to the median LOS with $\log N_{\rm H}=20.3\pm 0.1$ ($1\,\sigma$ confidence level) obtained
considering constant metallicity ($Z=0.3\pm 0.1\,\Zsun$) in absorbers with $100<\Delta<1000$. }
\label{F2}
\end{figure}

\section{Lower envelope and mean of the column density distribution with redshift} \label{sec:outliers}

The total sample of objects comprises 711 sources spanning the redshift range 0.01--8.2. Of these, 222 yield positive detections of 
an intrinsic column density. We approached this distribution with two main drivers.
The first is to find out the minimum column density allowed by all these data. 
This minimum $N_{\rm H}$ can be derived as the envelope
obtained taking a local absorber and evolve it at different redshifts.
In this case we are not interested in a fit of the different column densities, which provide a very biased sample (e.g. due to different selection 
criteria and intrinsic column densities, etc.), but a lower universal envelope applicable to all X--ray sources.
It has been shown that a local absorbed $N_{\rm H}(z=0)$ scales with redshift as $(1+z)^{2.4}$ (Campana et al. 2014). 
This means the effect of an absorber a $z=\bar{z}$ on the X--ray spectrum of a distant source scales as
$N_{\rm H}(z=0)=N_{\rm H}(z=\bar{z})/(1+\bar{z})^{2.4}$. The scaling index has been shown to slightly depend on the 
observing instrument with a small spread of $\pm0.05$ around the theoretical value of $2.40$ (Campana et al. 2014). 
This  minimally impacts the $N_{\rm H}$ evaluation.
To evaluate the minimum  column density, we consider a cold absorber (i.e. with the highest cross section) 
with present-day column density $N_{\rm H}(z=0)$ and use a value
larger by a factor $(1+z)^{2.4}$ at any $z>0$.
After a careful analysis of a few outliers in the $N_{\rm H}-z$ plane (see above), we fit the data with the requirement to leave below the curve
$1\%$ of all the data (mainly provided by upper limits). Upper limits are treated here as detections.
We find a value of $\log N_{\rm H}(z=0) \simeq 19.5$ (we express column densities in units of ${\rm cm^{-2}}$, see Fig. 1).

Concerning the evaluation of the mean contribution along the line of sight we adopted a different approach.
We stick to a well-selected and characterised sample and with high statistics ($>1,800$ counts per spectrum),
which is the AGN-E13 sample (Eitan \& Behar 2013).
In this sample only $43\%$ of the quasars show significant absorption. 
The authors tried to find which physical parameters of the quasars might drive the observed absorption, e.g. 
time (i.e. evolution), radio luminosity, or the X--ray luminosity. The most likely interpretation is in terms of an increasing 
absorption along the line of sight, possible with  minimal contamination from intrinsic absorption in the AGN themselves.
We note in fact that there are no highly extinguished objects in the sample, hinting at the lack of Type II, intrinsically heavily absorbed, AGN.

We made minimal changes to the AGN-E13 sample, simply getting rid of the two outliers above. The sample then consists of 56 objects. 
More than half of the $N_{\rm H}$ estimates are upper limits (31, $55\%$). This large fraction of upper limits calls for censored statistics. 
It is not easy to perform a linear regression test, keeping fixed the value of the slope (to 2.4). For this reason we rectified all
the $N_{\rm H}$ estimates by dividing them for the term $(1+z)^{2.4}$. We then approached the new distribution, searching for the 
best (censored) constant fit. 
We used programs of the ASURV suite (Isobe et al. 1986). 
In ASURV we made use of the Kaplan-Meier estimator.
This censored fit results in a mean column density absorption of $\log N_{\rm H}(z=0)=20.3\pm0.1$ ($1\,\sigma$ confidence level). 

If we restrict ourselves to a smaller sample, obtained by requiring at least 5,000 counts, we retain 24 objects, with only 5 upper limits ($21\%$).
In this case where the impact of upper limits is much lower, we obtained a minimum column density of $\simeq 19.3$ 
and a mean column density absorption of $\simeq 20.4$.

\section{Cosmological simulations} \label{sec:simule}

We have used the simulation presented in Pallottini et al. (2013), 
whose main features are recalled below.
The simulation is performed using the publicly available adaptive mesh refinement
code {\tt RAMSES} (Teyssier 2002), which can be described as a fully threaded tree data structure where the 
hydrodynamical adaptive mesh refinement scheme is coupled with a particle mesh N-body solver employing a 
cloud-in-cell interpolation scheme to solve the Poisson equation.

Note that heating-cooling processes in {\tt
RAMSES} are already implemented (Theuns et al. 1998; Aubert et al. 2008), including the
treatment of an external  UV ionising background (Haardt \& Madau 1996);
we have not included star formation, metal enrichment, nor internal radiative feedback for the following reasons. 
The treatment of winds, metal transport, and mixing remains an extremely difficult challenge for all current simulations. 
Hence we have decided to use a simulation containing only physics that we can reasonably keep under control (gravitation, 
structure formation shocks and radiative processes, adiabatic cooling), and combined it with different prescriptions for the 
metal spreading taken from specific studies or proposals by other leading groups. This appears reasonable as our aim is to 
bracket the uncertainties of the problem rather than constraining different feedback and enrichment models.

The simulation box size of $100\,h^{-1}$ ($h$ is the normalised Hubble constant in units 
of 100 km s$^{-1}$ Mpc$^{-1}$) comoving Mpc is evolved from $z=100$ to
$z=0$ using $256^{3}$ dark matter (DM) particles and a corresponding number of base
baryonic grid; we allow six additional levels of refinement adaptively chosen with a
Lagrangian mass threshold-based criterion. This yields a mass resolution of $1.65
\times 10^{10}\ \Omega_{\rm dm}\,h^{-1}\ M_{\odot}$ for the DM and a formal spatial
resolution of $6\,h^{-1}$ kpc for baryons. We have assumed a WMAP7
cosmology (Larson et al. 2011) with $\Omega_{\Lambda}= 0.727$, $\Omega_{\rm dm}=
0.228$, $\Omega_{\rm b}= 0.045$, $n=0.967$, $\sigma_{8}=0.811,$ and $h=0.704$.

Simulation results, in particular the baryon thermodynamic, are consistent with numerical studies (Rasera et al. 2006; Peeples et al. 2010).  
For example, Peeples et al. (2010) simulated a 
$12.5 h^{-1}$ Mpc comoving box with a high ($2\times288^{3}$ particles) and low ($2\times144^{3}$ particles) 
resolution using the smooth particle hydrodynamic (SPH) code {\tt GADGET-2} (Springel 2005). 
At $z=3,$ Peeples et al.  found an equation of state very similar to  that obtained here.
Additionally, when restricting to the quasi-linear regime, baryons evolution is in accordance with semi-analytical 
models (Miralda-Escud\'e, Haehnelt \& Rees 2000; Choudhury, Padmanabhan \& Srianand 2001).

For the present work we have extracted various LOS from the light cone going from
$z=0$ to $z=2$. As in Pallottini et al. (2013), the LOS are randomly selected from
contiguous snapshots of the simulation. For this work a suitable convergence of the
results is obtained with $N=100$ LOS.

\begin{figure*}[!th]
\centerline{
\includegraphics[width=1.0\textwidth]{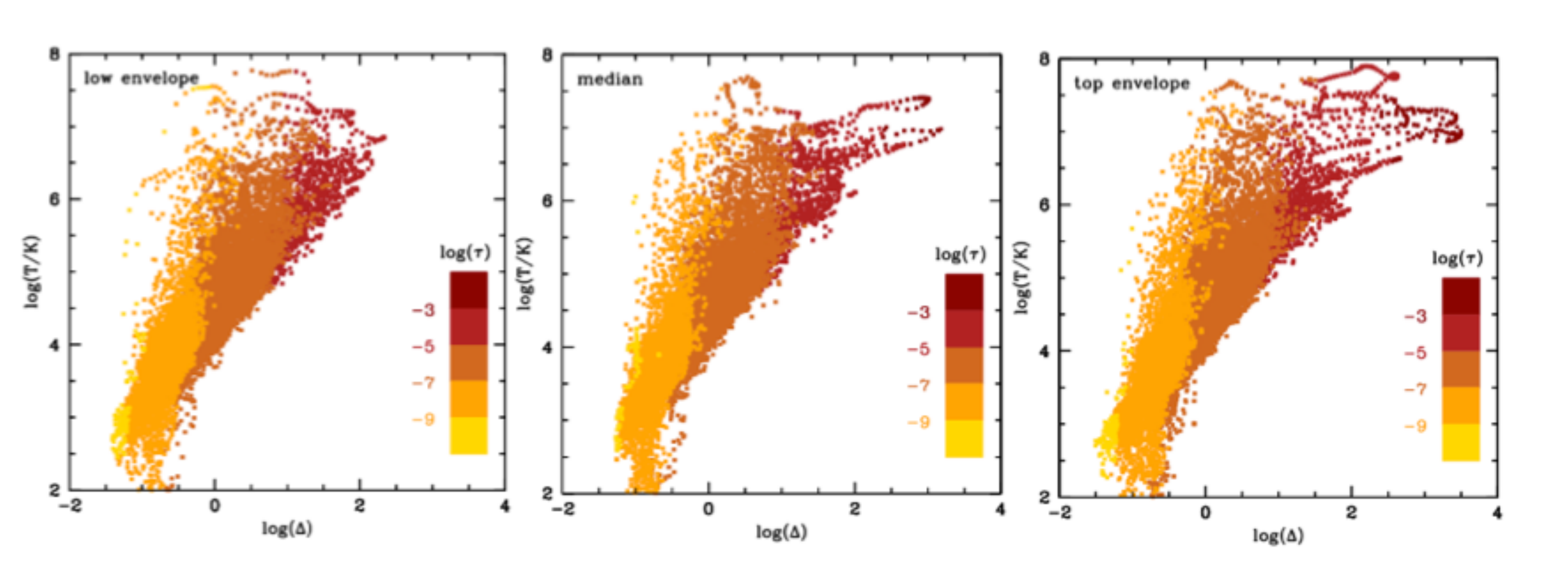}
}
\caption{
Individual contribution of cells in the $\Delta_i-T_i$ plane for the
reference momentum-driven wind model along three different LOS: {\it
Left panel a)} refers to the LOS with the lowest absorption. The optical depth 
along this LOS in our simulation is $\tau(\rm {0.5 keV})=0.06$.  {\it Central
panel b)} indicates the median LOS corresponding to the solid, black line in
Fig. 2, the median optical depth is $\tau(\rm {0.5 keV})=0.21$.
The {\it right panel c)} refers to the LOS defining the top envelope
of simulated curves in Fig. 2. In this case the optical depth is $\tau(\rm {0.5 keV})=0.88$.
The colour scale represents the value of
$\tau_i$ computed at 0.5 keV. The LOS with the lower absorption lacks of
cells with $\Delta_i>100$. Instead these cells dominate the X--ray
optical depth in the median and top LOS, where most ($\sim 60\%$) of the
absorption is provided by two systems with $\Delta_i>300$ and $T_i>10^6$ K (dark red points).
}
\label{F3}
\end{figure*}

\section{Cosmological absorption} \label{sec:tau}

We compute the X--ray optical depth for photons seen by the observers at the energy $E$ by summing up the contribution of each 
cell along the LOS up to the source redshift. The $i$-th contribution at the energy $E^\prime=E(1+z_i)$ is given by  
$\tau_i(E^\prime)=Z_i N_{{\rm H}_i} \sigma(E^\prime,T_i,\xi_i)$, where $N_{{\rm H}_i}$, $T_i$ are the gas column density and 
temperature of the $i$-th cell and are provided directly by the simulation. The ionisation parameter is $\xi_i\simeq 4\pi F_{CXRB}/n_{e_i}$, 
where the integrated cosmic X--ray background spectrum is $F_{CXRB}\sim 2.9\times 10^{-7}$ erg cm$^{-2}$ s$^{-1}$ sr$^{-1}$ in the 
energy band 0.005--300 keV (Deluca \& Molendi 2004)
and the election density of the $i$-th cell, $n_{e_i}$, is a factor $\sim 1.2$ higher 
than the hydrogen density in the cell. Finally, we computed the photo-electric cross-section $\sigma$ using the XSPEC model 
{\tt ABSORI} (Done et al. 1992) 
for the temperature $T_i$ and ionisation parameter $\xi_i$ , assuming the solar composition on a parameter grid.
We assigned the metallicity of the $i$-th cell, $Z_i$,  according to different assumption and physical motivated models (see next Section).
Therefore,  we consider  the temperature and ionisation state for each cell, depending on the density, and we work out the 
appropriate cross section. This is computed at 20 different ($z=0$) energies logarithmically spaced in the 
0.3--10 keV energy range. All these contributions are then summed up at the given redshift to provide the total absorbing column 
density along the line of sight. 
To compute the intrinsic $N_{\rm H}$ to be compared with observational data, we fit the resulting cross section with a cold absorber 
with $\xi=0$ at the redshift of the background source. 

\section{Metal enrichment models}

The simulation described in Sect. 4 does not directly provide  the metallicity of  the gas that is needed to compute the X--ray 
optical depth. Metals produced by supernova and during the asymptotic giant branch phase are expected to be spread in the circumgalactic 
and intergalactic medium by galactic outflows (Oppenheimer \& Dav\'e 2008; Kobayashi, Springel \& White 2007). 
In the following we will describe how we assign metallicity to each simulated cell based on its over-density ($\Delta -Z$ prescription) and 
how our results depend on the different choices. In particular, we provide three different descriptions for the metal enrichment of
increasing complexity, starting from back-of-the-envelope calculations to physically motivated models. The estimates we derive are consistent.

\subsection{Simple estimates}

The photo-electric optical depth (opacity) of a homogeneous IGM of constant metallicity $Z$ (in solar units) at observed photon 
energy $E$ and temperature $T$ can be expressed as (Behar et al. 2011; Starling et al. 2013): 
$$
\tau_{IGM}(E,\,z,\,Z)=\int_{0}^{z} n_{\rm H}(z')\,\sigma(E,\,z',\,Z,\,T)\,c\,{{dt'}\over{dz'}} \, dz'. \ \eqno   (1)
$$
The absorption cross section $\sigma(E,\,z',\,Z,\,T)$ for a solar mixture of elements scales linearly with $Z$.
For a homogeneous and isothermal IGM one can write the above equation as
\setcounter{equation}{1}
\begin{eqnarray}
\tau_{IGM}(E,\,z,\,Z)={{n_0\,c}\over{H_0}} && \int_{0}^{z} Z \,\sigma(E,\,z',\Zsun,\,T)  \nonumber \\
&\times & {{(1+z')^2}\over{\sqrt{(1+z')^3
\Omega_{\rm M}+\Omega_{\Lambda}}}} \, dz'   
,\end{eqnarray}
where $n_0=1.7\times10^{-7}$ cm$^{-3}$ is the present-day mean IGM density, $H_0 =71$ km s$^{-1}$ Mpc$^{-1}$ is 
the Hubble constant, $\Omega_{\rm M}=0.27$ and $\Omega_{\Lambda}=0.73$ are, respectively, the present-day matter and 
dark energy fractions of the critical energy density of the Universe, and $c$ is the speed of light.
We selected an energy of $E=0.5$ keV just below the oxygen edge. 
On the right hand side of Eq. 2, $\tau$ can be expressed as $\tau=N_{\rm H}\sigma$. To obtain the IGM metallicity, 
we can consider our low column density envelope $\log{N_{\rm H}}=19.5$. In this case, however, the appropriate cross section to use 
on the right-hand side of Eq. 2 is the one pertaining to a cold absorbing material since all the column densities were obtained 
in this way, i.e. $\sigma=\sigma(E=0.5{\rm \,keV},\,z',\Zsun,\,\,T=0)$. We can then select a single temperature for the IGM and work out the 
cross section dependence on redshift using the {\tt ABSORI} model within XSPEC, as described above (this dependence is 
not trivial and cannot be expressed as a simple power law). 
Based on our minimum absorbing column density, working out the integral (from $z=0$ to up to $z=2$, 
it being negligible above this value) and further assuming an 
IGM $T=10^6$ K or $T=10^5$ K, we derive a mean metallicity  of $Z=0.043\,Z_\odot$ or $0.025\,Z_\odot$, respectively.
For a neutral (cold) IGM the  cross section on the right- and left-hand sides of the equation can be simplified, and 
we derive $0.014\,Z_\odot$. 

\subsection{Simple models}

To bracket all possible enriched scenarios, we consider two limiting cases: (i) constant metallicity, i.e. $Z_i/\Zsun= Z_a=$ const. in all 
cells, and (ii) a no-wind model. In the latter case, $\log(Z_i)= \log(Z_b)+\log(\Delta_i)$, with $Z_b=$const. For both scenarios we compute 
the minimum envelope 
curve by considering the LOS with the lowest absorption (we checked that we obtain consistent results using the lowest three LOS).
The values of our free parameters ($Z_a$ and $Z_b$, respectively) are then 
obtained by requiring that only $1$\% of the data points fall below the simulated curve. We find: (i) $Z_a=0.030\,\Zsun$ ; and 
(ii) $Z_b=0.0018\,\Zsun$, the two curves being almost identical in Fig. 2.
From this we estimate the metal cosmic density by eq. (1) of Ferrara et al. (2005)
to be $\Omega_{\rm Z}^{IGM}=1.77\times10^{-5}$ and $1.17\times10^{-5}$, respectively. In spite of the very different (extreme) 
assumptions, the two models result in similar metal column and cosmic densities.

In a similar manner, we compute the median of the simulated LOS distribution for the constant metallicity case. 
The signal is in this case dominated by over-densities in 
the range 100--1000. To match the observed median found in Sect. 3, a mean metallicity of $Z=0.3\pm 0.1\,\Zsun$ 
associated with these over-densities is required.

It is meaningful to compare the above numbers with the comic density of metals produced by SNe by $z=1$ where the peak of the cosmic 
star formation rate is expected. Therefore, we integrate the star formation history (SFH) as computed by Li (2008) from high redshift ($z
\sim 15$, well enough to assure convergence)
down to $z=1, $ assuming that the specific fraction of heavy elements produced by stars is 0.03 (Peeples et al. 2014).
Uncertainties on the integrated SFH should be minor since it rapidly decreases at $z>2$ and Pop III gives 
a negligible contribution for $z<4$ (Tornatore et al. 2007; Pallottini et al. 2014). Uncertainties in the nucleosynthetic yields are 
instead larger (and not easy to quantify, e.g. Wilkins et al. 2008) and we decided to stick to the Peeples et al. (2014) estimate
(also for comparison purposes).
We obtained $\Omega_{\rm Z}(z=1)\sim 1.4\times 10^{-4}$. Thus, we find that 12\% and 8\% of the metals 
produced by SNe by $z=1$ are expelled far from galaxies in regions with $\Delta<100$ for model (i) and (ii), respectively.

\begin{figure*}[!t]
\centerline{
\includegraphics[width=0.5\textwidth]{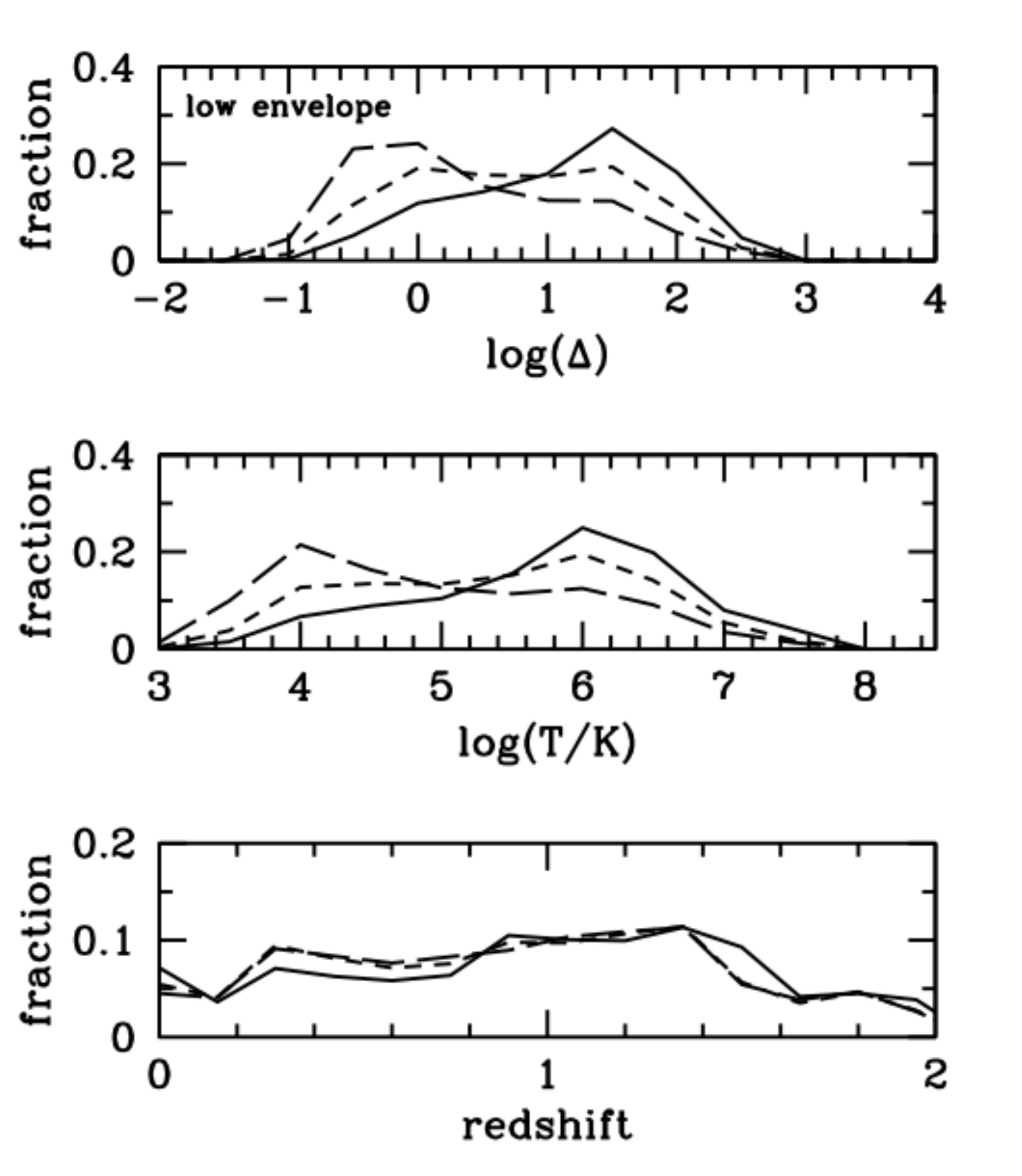}
\includegraphics[width=0.5\textwidth]{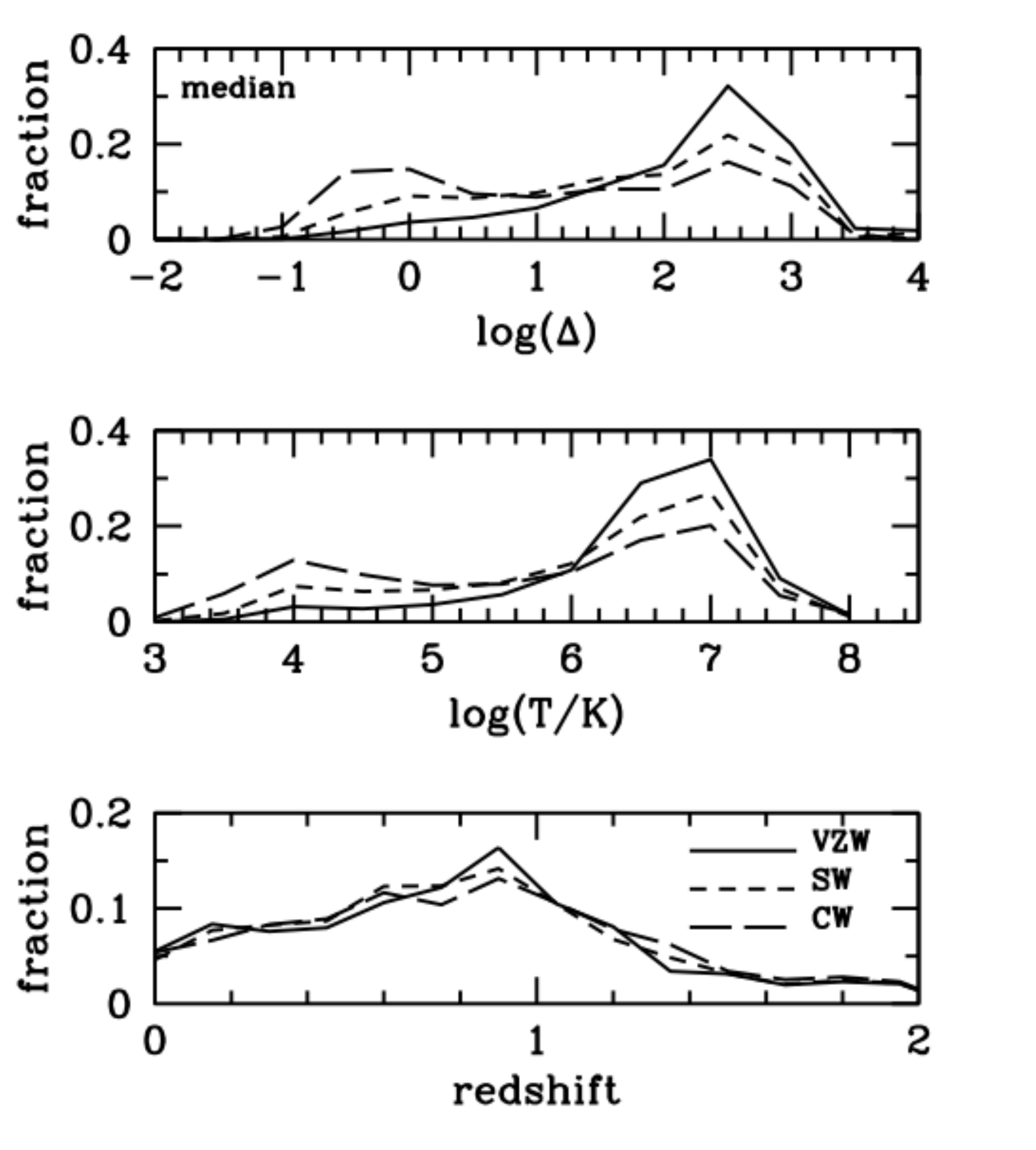}
}
\caption{
Fraction of the X--ray optical depth at the observed energy $E=0.5$ keV 
as a function of the cell over-density, temperature, and redshift (from 
top to bottom). The left panels refer to the mean of the five LOS with 
the minimal absorption, while the right panels refer to the mean of $68\%$ of the 
LOS around the median. Solid, short-dashed, and long-dashed lines refer 
to the VZW, SW, and CW models, respectively. Note that the CW model slightly
overpredicts the minimum envelope of the $N_{\rm H}-z$ distribution, as noted in the text.}
\label{F4}
\end{figure*}

\subsection{Physically-motivated enrichment models}

We assign the metallicity to each cell in our simulation by adopting the distributions of metallicities as a function of both over-density and 
redshift predicted by three different wind models of Oppenheimer et al. (2012): 
(a) a constant wind (CW) model where the ejection velocity 
is $v_{\rm wind}=680$ km s$^{-1}$ and mass loading factor of $\eta=2$ for all galaxies; (b) a slow wind (SW) model in which the ejection velocity 
is half of the previous velocity; (c) a momentum-conserved (VZW) model with $v_{\rm wind}=3\sigma_g\sqrt{f_L-1}$ in which $f_L$ is the luminosity in 
units of the Eddington luminosity required to expel gas from the galaxy potential, $\eta=\sigma_0/\sigma_g$, and $\sigma_0=150$ km s$^{-1}$ 
, and $\sigma_g$ is the galaxy's internal velocity dispersion (see Oppenheimer et al. 2012 and Oppenheimer \& Dav\'e 2008 for model details). 
Oppenheimer et al. (2012)
show that in spite of the significant differences among the metal enrichment prescriptions, all wind models produce 
metal-line statistics within a factor of two of existing HST/COS observations (Danforth et al. 2014; Liang \& Chen 2014;  Stocke et al. 2013;
Tilton et al. 2012;  Prochaska et al. 2011), with the physically motivated VZW model preferred by the data. 
In the following we will refer to the VZW model as our reference prescription.

For each wind model we compute the expected $N_{\rm H}-z$ distribution following the procedure described above. Note that since the 
metallicity is fixed, the model has no free parameters. In Fig. 2 we show the results for our reference prescription. 
In the following we will discuss the main results obtained with this metal enrichment recipe and the differences with the alternative wind models.

In the reference prescription (momentum-driven winds) most of the absorption comes from regions with $10<\Delta<100$ 
enriched to $Z\simeq 0.030\,\Zsun$ (see Fig. 3). 
They can be naturally identified with the genuine IGM. These regions contain $6\%$ of the metals produced by 
SNe by $z=1$; an additional $4\%$ has been transported into regions of even lower over-density. 

The {\it  minimum envelope} of the $N_{\rm H}-z$ distribution of our reference prescription  reproduces very well the observations with $<1$\% 
of data falling below the lowest simulated curve. The metal cosmic density derived from this LOS is $\Omega_{\rm Z}(z=1)=1.16\times 10^{-5}$, i.e. 
$\sim 10$\% of the metals produced by SNe by $z=1$.  This is not true for alternative models: $\sim 3$\% of data are found to be outliers in the 
CW model, while this fraction reduces to $\sim 1$\% in the SW model. This reflects the fact that metal enrichment of the regions at the lowest over-densities 
is much faster and efficient in the CW model with respect to the VZW model. In particular, the CW model predicts a ten times larger metallicity in 
$\log(\Delta)=0$ regions than the VZW model. This is further confirmed by looking to the fraction of the X--ray optical depth provided by the cell
at different over-densities shown in the right panel of Fig. 4 (red lines). In the CW model most of the signal is provided by regions 
with $\log(\Delta)\sim 0$ while in the VZW model regions with $1\le\log(\Delta)\le 2$ dominate. The distribution of the SW model is 
intermediate between the two. This evidence is confirmed by looking at the distributions in terms of the gas temperature (right panel of 
Fig. 4) with cold regions dominating in the CW model. 

These differences become less important when looking at the {\it \textup{median}} of the X--ray absorption distribution (black lines in Fig. 4). 
Indeed, the median of the distribution is only a little sensitive to the assumed model. In all models the signal is dominated by warm-hot 
($6.5\le\log(T/{\rm K})\le 7.5$), collapsed ($\log(\Delta)>2$) regions. Most of the absorption arises from distant intervening sources at $z\ge 0.5$. 
The resulting $N_{\rm H}-z$ curves of different models lie within a factor of 1.3 from our reference prescription.

\section{Discussion and conclusions}

Although sensitive to  a wide range of metal ionisation states and species, X--ray experiments, because of limited spectral resolution, 
lack the ability of optical/UV studies to identify individual absorbers along the LOS. Rather, X--ray absorption gives a cumulative 
view of metals in the $z\sim 0-2$ redshift range. 
We explore the possible existence of a minimum X--ray absorption contribution due to diffuse metals. 
Data support the existence of this 'metal fog', even if a direct measurement is prevented by current (and future) observational limits.
In addition, if background X--ray sources are projected on top of denser intervening regions larger column densities are retained, resulting 
in nearly an order of magnitude spread in X--ray column densities.

A simple estimate of the metal content in the ideal case of a homogeneous, 
isothermal IGM with mean metallicity $Z$ can be obtained by comparing the 
photo-electric optical depth with the observed  minimum absorbing column density.
For $T=10^6$ K ($10^5$ K) we derive, by integrating in $0<z<2$,  $Z=0.043\,Z_\odot$ ($Z=0.025\,Z_\odot$), 
corresponding to a metal cosmic density of metals in the IGM of $\Omega_{\rm Z}^{IGM}=2.4\times10^{-5}$ ($\Omega_{\rm Z}^{IGM}=1.4\times 10^{-5}$).  
For a neutral (cold) IGM we instead derive  $Z=0.014\,Z_\odot$.
To provide a more complete description of the absorbing medium we rely on extensive cosmological 
simulations. To compute the metal column density of the 
intervening material, we consider 100 LOS to distant sources through a $100\, h^{-1}$ comoving Mpc 
Adaptive Mesh Refinement cosmological simulation (Pallottini et al. 2013).

We compute the X--ray optical depth for photons seen by the observer at energy $E$ as 
$\tau(E) \simeq \sum_i Z_i N_{{\rm H}_i} \sigma[E(1+z),T_i,\xi_i]$, where $N_{{\rm H}_i}$, $\sigma$, $\xi_i$,  $T_i$ , and $Z_i$ are the gas column density, 
the total cross section, ionisation parameter (set by the cosmic X--ray background), temperature, and metallicity in the $i$-th simulation cell, respectively. 
All these quantities are drawn from the simulation box, except for the metallicity $Z_i$. 
As considerable uncertainty in numerical simulations on the transport and mixing of heavy elements still remains,
we consider two limiting cases to bracket all possible enrichment scenarios: 
(i) constant metallicity, i.e. $Z_i/\Zsun={\rm const.}$ in all cells along the LOS;  (ii) a no-wind model in which metals remain bound 
to their formation sites. 

The {\it minimum envelope} of the $N_{\rm H}-z$ distribution should be provided by the simulated LOS with the lowest absorption. 
Based on our simulations, this LOS does not contain any absorber with over-density $\Delta_i \equiv \rho_i / \bar\rho >100$,
where $\rho_i$ is the gas density in the cell and $\bar\rho$ the mean cosmic gas density. These regions are hot 
$T\simeq 10^{5-7}$ K and far from collapsed structures.
In order to match the minimum envelope of the $N_{\rm H}-z$ distribution, we find that $8-12\%$ 
of the metals produced by SNe by $z=1$ must have been ejected far from galaxies in the IGM.
This result is almost insensitive to the distribution of metals in regions with different over-densities. 
The mean metallicity of the IGM is $Z=0.03\,Z_\odot$.
More realistic prescriptions (Oppenheimer et al. 2012) provide results within the above range (see above).
Traces of these structures can be found through detailed X--ray absorption studies (Nicastro et al. 2013).
Indications of a similar metal content in the IGM came very recently from UV studies (Shull, Danforth \& Tilton 2014).

The {\it mean} of the X--ray absorption distribution based on AGN is even more telling. In the simulations the median LOS is 
dominated by 2--5 absorption systems with large over-densities ($\Delta>300$) at $z\sim0.5-1.2$.
The total column density contribution of these regions (at their respective redshift) is $\log N_{\rm H} \simeq 21.5$. 
The absorbing systems have characteristic projected sizes of 0.1--1 Mpc and are associated with dark matter halos of virial mass 
$\sim (0.2-2)\times 10^{13}\ M_\odot$, corresponding to a virial temperature of $\sim  (3-15)\times 10^6$ K.
These structures are about ten times more massive than the Milky Way, suggesting that the absorption is due to the 
extended circumgalactic reservoir of gas of (small) galaxy groups rather than to single isolated galaxies. 
Given that the mean metallicity in these regions ($\Delta\sim 100-1000$) is $0.3\pm0.1\,Z_\odot$ (in agreement with 
current observations Humphrey et al. 2012 and Su 2013), we can conclude that they contribute an additional $\sim 11\pm4\%$ to the metal budget. 

Direct detection of this hot, enriched material in the X--ray band out to the virial radius (defined such that the mean density of a halo 
of mass $M$ is 200 times the cosmic mean) of galaxies is challenging with current X--ray facilities (Sun et al. 2009). 
However, the presence of these reservoirs might have been identified by recent optical/UV
absorption studies, thus paving the road to deeper, dedicated studies (Mulchaey et al. 1996; Stocke et al. 2013).

\begin{acknowledgements}
We acknowledges useful discussions with E. Feigelson, F. Gastaldello, A. Moretti, \& L. Sage.
\end{acknowledgements}



\appendix
\section{Outliers} \label{sec:outliers}
 
Based on Fig. 1, we identified 11 main outliers in the distribution of intrinsic column densities ($N_{\rm H}(z)$). 
If this is a true cosmological effect no outlier should be present. Therefore we carefully study these outliers in order to find an explanation
for their low intrinsic column densities. We approached the X--ray spectra of these sources and tried to verify if the 
evaluation of the column density is completely satisfactory from a statistical point of view.
In this respect we disregard AGN with an intrinsic column density $\log N_{\rm H}<19$ and $z<0.3$ because of a too small intervening length scale 
and, more importantly, because of the AGN influence in cleaning a substantial fraction of the surrounding environment. Five AGN were discarded in this way.
For the other six AGN, we provide a detailed description.


\subsection{QSO B0909+5312 at $z$=1.377.}

QSO B0909+5312 was observed by \xmm\ in May 2003. Data were retrieved from the \xmm\ archive 
and reprocessed using SAS 13.5.0. Useful \xmm\ data were taken from the entire 13 ks and 17 ks exposures of pn and MOS detectors, respectively. 
Data were extracted following common practice and spectrally binned to 50 counts per energy channel; {\tt arf} and {\tt rmf} files were generated 
using SAS tasks. Fitted energy ranges are 0.2--10 keV  and 0.3--10 keV for the pn and MOS instruments, respectively.
Eitan \& Behar (2013) fitted the same data with an absorbed power-law model. They adopted a Galactic column density of 
$N_{\rm H}^{\rm Gal}=1.49\times 10^{20}$ cm$^{-2}$ (Kalberla et al. 2005) 
plus and intrinsic column density at the AGN redshift. Based on their spectrum they derive an upper limit of $N_{\rm H}(z)<7\times 10^{19}$ cm$^{-2}$.
We fit our data with a higher binning  to increase statistics. We adopted the {\tt TBABS} model within XSPEC to account for X--ray absorption 
(Wilms, Allen \& McCray 2000) 
and with the {\tt wilm} solar abundance pattern (this is very similar to Asplund et al. (2009) and in the case of oxygen, the main 
driver for the absorption, is exactly the same) and the {\tt vern} photoelectric absorption cross sections. 
We obtain similar results ($N_{\rm H}(z)<5\times 10^{19}$ cm$^{-2}$, $90\%$ confidence 
level for one parameter of interest, i.e. $\Delta\chi^2=2.71$) but with a reduced $\chi^2_{\rm red}=1.14$ (for 187 degrees of freedom, dof). 
Fitting the same data with a broken power law we obtain instead $\chi^2_{\rm red}=0.77$ (185 dof).
An F-test indicates that the addition of the spectral break is significant at the $>8\,\sigma$ confidence level.
The intrinsic column density in this case is $N_{\rm H}(z)<5\times 10^{20}$ cm$^{-2}$, perfectly consistent with our lower envelope.

\subsection{Two AGNs in the Lockman hole at $z$=1.204 and $z$=1.113.}

J105239.6+572432 (J1052 at $z=1.113$) and J105316.9+573552 (J1053 at $z=1.204$) are two Type I AGN in the Lockman hole 
with a very low absorption ($N_{\rm H}(z)<6\times 10^{19}$ cm$^{-2}$ and $N_{\rm H}(z)<9\times 10^{19}$ cm$^{-2}$, respectively (Corral et al. 2011).
We took the longest \xmm\ observation of the Lockman hole not affected by soft proton flare covering 100 ks. 
Data were extracted as above and binned to 25 counts per spectral bin. 

J1052 can be fit with a simple power law, resulting in a $\chi^2_{\rm red}=1.15$ (288 dof) and $N_{\rm H}(z)<3\times 10^{20}$ cm$^{-2}$ 
($N_{\rm H}^{\rm Gal}=5.5\times 10^{19}$ cm$^{-2}$). 
A better fit is obtained with a broken power law, with  $\chi^2_{\rm red}=0.92$ (286 dof) and $N_{\rm H}(z)<6\times 10^{20}$ cm$^{-2}$. 
An F-test indicates that the addition of the spectral break is significant at $>8\,\sigma$.

J1053 falls on the dead CCD in MOS1 and across dead columns on the pn detector. The fit with a single power law is good 
($\chi^2_{\rm red}=1.01$, 129 dof) and returns $N_{\rm H}(z)<2\times 10^{20}$ cm$^{-2}$. 

Both values are consistent with our lower envelope.

\subsection{QSO B1345+584 at $z$=0.646.}

QSO B1345+584 is a Type I AGN. 
Corral et al. (2011) derived an upper limit on the intrinsic column density of $N_{\rm H}(z)<1\times 10^{19}$ cm$^{-2}$ for this source.
It was observed by \xmm\ in Jun 2001 for 40 ks, even if it was not on-axis. 
Because of a strong proton flare activity at the end of the observation only
the first 25 ks and 29 ks were retained for the pn and MOS2 instruments, respectively (the source falls on a CCD gap in MOS1).
Data were extracted as above and binned to 25 counts per spectral bin. 
The fit with a power law  (assuming $N_{\rm H}^{\rm Gal}=1.1\times 10^{20}$ cm$^{-2}$) provides a $\chi^2_{\rm red}=1.19$ (308 dof) with a clear 
under prediction of the high energy spectral part. In this case the intrinsic column density is negligible ($N_{\rm H}(z)<2\times 10^{19}$ cm$^{-2}$).
Fitting the data with a broken power law ,we obtain $\chi^2_{\rm red}=0.98$ (306 dof). An F-test indicates that the addition of the spectral break is 
significant at $>8\,\sigma$.
The intrinsic column density in this case is $N_{\rm H}(z)<7\times 10^{19}$ cm$^{-2}$, in line with similar upper limits at this redshift.

\subsection{QSO B1157--1942 at $z$=0.450.}

QSO B1157--1942 is the  lowest redshift quasar ($z=0.45$) in the AGN-E13  sample. They quote a very low intrinsic column density
of $N_{\rm H}(z)<1\times 10^{19}$ cm$^{-2}$. 
\xmm\ observed the quasar on Jun 2008. The exposure time is 25 ks and 24 ks for the MOS and pn detectors, respectively.
Data were extracted as above and binned to 50 counts per spectral bin. Fitting the data with an absorbed power -law model ($N_{\rm H}^{\rm Gal}=3.2\times 10^{20}$ 
cm$^{-2}$) , we derive $N_{\rm H}(z)<6\times 10^{18}$ cm$^{-2}$ (in line with previous results) and $\chi^2_{\rm red}=1.16$ (791 dof). The residuals show clear
signs of an underlying, redshifted iron line. Adding a free Gaussian component at the quasar redshift, the fit improves to $\chi^2_{\rm red}=1.02$ (788 dof,
F-test probability $>8\,\sigma$). The derived limit on the intrinsic column density $N_{\rm H}(z)<1\times 10^{20}$ cm$^{-2}$ is in line with our expectations.

\subsection{2E 1852 at $z$=0.312.}

This Type I AGN is reported to have an intrinsic column density $N_{\rm H}(z)<1\times 10^{19}$ cm$^{-2}$ (Corral et al. 2011).
The object was observed by \xmm\ on Apr 2000 in small window with the two MOS and in full window with the pn.
Some flaring activity was present and filtered out reducing the exposure time to 54 ks and 22 ks for the MOS and pn, respectively.
MOS data were fitted in the 0.6--10 keV range to encompass calibration uncertainties in small window mode. All data were binned to have at least 25 counts
per spectral bin.
We fit the spectra with an absorbed power-law model ($N_{\rm H}^{\rm Gal}=3.3\times 10^{20}$ cm$^{-2}$). We derive a good fit 
with $\chi^2_{\rm red}=1.07$ (1316 dof) and $N_{\rm H}(z)=(4.3\pm0.3)\times 10^{20}$ cm$^{-2}$, much larger than the reported value.

\end{document}